\documentclass{elsart}
\usepackage{graphics}
\usepackage{graphicx}
\usepackage{epsfig}
\usepackage{amssymb,amsmath}

\begin{document}
\begin{frontmatter}
\title{Wavelet-Based Prediction for Governance,
Diversification and Value Creation Variables}
\author{Ines Kahloul}
\address{Higher Institute of Management, Department of Finance,\\
41, Rue de la Libert\'e, Bouchoucha - 2000 Bardo, Tunisia.}
\author{Anouar Ben Mabrouk}
\address{Computational Mathematical Laboratory, Department of Mathematics,\\ Faculty of Sciences,
5000 Monastir, Tunisia.} \ead{Corresponding Author:
anouar.benmabrouk@issatso.rnu.tn}
\author{Slaheddine Hallara}
\address{Higher Institute of Management, Department of Finance,\\
41, Rue de la Libert\'e, Bouchoucha - 2000 Bardo, Tunisia.}
\begin{abstract}
\small{We study the possibility of completing data bases of a
sample of governance, diversification and value creation variables
by providing a well adapted method to reconstruct the missing
parts in order to obtain a complete sample to be applied for
testing the ownership-structure/diversification relationship. It
consists of a dynamic procedure based on wavelets. A comparison
with Neural Networks, the most used method, is provided to prove
the efficiency of the here-developed one. The empirical tests are
conducted on a set of French firms.}
\end{abstract}
\end{frontmatter}
\maketitle
\section{Introduction}
The relationship relationship ownership-structure/diversification
is of great interest in financial studies due to the role of
diversification as a strategic choice affecting the capacity of
firm development during long periods, its financing needs, its
performance as well as related risks. Such a decision may cause
some inter-actors interest conflicts and it may be imposed within
its actionnarial structure. The main problem in studying the
relationship ownership-structure/diversification is the necessity
of conducting it on a complete data basis. So, with missing data,
the study can not be well conducted and no conclusions can be
pointed out correctly. This leads to a second problem consisting
of completing data bases by reconstructing missing parts which
induces also the third most difficult problem based on the fact
that governance, diversification and value creation variables are
short time series (One value on a year) leading any reconstruction
to be difficult. This motivates our work here consisting of
providing a well adapted method to reconstruct the missing parts
on a sample of governance, diversification and value creation
variables in order to obtain a complete sample to be applied for
testing the ownership-structure/diversification relationship.

It is well known and often obvious that statistical or empirical
tests are well conducted with complete sample datum. But, in many
major cases, it may happen that some data are missing or can not
be provided by the owner especially when dealing with fuzzy data,
secured or coded data, etc. In these cases, the non availability
may yield essential problems effecting the conclusions that can be
conducted from empirical tests. Indeed, the incomplete file
treatment places the statistician in front of difficulties in
applying theoretical concepts. Furthermore, it can not generalize
the discovered conclusions on all the population especially when
dealing with important missing data parts. Otherwise, in multi
regression analysis, some variables are non informative or
redundant. Besides, the absence of data can eliminate the
incomplete observations. The loss of information thus gotten can
be considerable if numerous variables present missing values on
different individuals. This problem therefore, of missing data
should be managed with precaution in order to avoid all slanted in
evaluating and analyzing results. The researches on this problem
are growing up and different methods have been investigated in
order to overcome the difficulties. From a practical point of
view, especially for financial analysts and/or economists, it
seems that the most effective idea is about the reconstruction of
such data to obtain quasi-complete samples. Different methods have
been proposed and tested, such as Neural Networks, Fuzzy Logic,
Fourier Analysis, Autoregressive Models, etc. See for intense
Miller (1990), Wodzisaw and (1996)), Karp et al (1988). However,
these methods have not been efficient in all cases and in contrast
they can lead to noised reconstructed data. Furthermore, it has
been proved that such methods can not analyze all types of data.
Let us comment them briefly and separately. Neural Networks
approach has been widely used as an alternative for signal
approximation since it provides a generic black-box representation
for implementing mappings from an inputs' space to an outputs'
one. The approximation accuracy depends on the measure of signal
closeness. Neural Networks implementation is essentially based on
key activation functions. So, the crucial point to obtain best
approximations turns around how one should chose these functions.
Almost all papers dealing with the subject are based on the
assumptions that such functions are integrable, sigmoidal
squashing, monotone, continuous and/or bounded. More precisely,
there are not many choices of such functions but these are limited
to sigmoid, tanh and Heaviside. Fuzzy Logic is based on two main
principles. First, it requires to fix the input rules and to
conduct data to satisfy or not these rules which is somehow
ambiguous. Autoregressive Models are also based on one main
hypothesis affirming that some dependence of present data on its
history should exist. Fourier Analysis which consists in
developing periodical phenomena in a superposition of oscillating
ones based on sine and cosine with known frequencies and
amplitudes, provides precious information on the analyzed signal.
For example, for higher frequencies, the signal varies slowly
conversely to low frequencies where it varies rapidly. But, the
major problem is the fact that Fourier Analysis can not localize
well the region of high or low variations. This lack in time and
frequency localization lets researchers to think about new tools
taking it into account. Recently, until the 80's, a new
mathematical tool has been introduced issued from signal
processing theory. It consists of Wavelet theory which has been
proved to be a best candidate that refines the existing methods to
be more adequate to data problems. The crucial idea in wavelet
analysis consists in decomposing a time series into independent
components with different scales well localized in time and
frequency domains. So, wavelet analysis is until its appearance
strongly related to scaling concept. Indeed, an analyzing basis is
always obtained from a source function called mother wavelet by
scaling and translation actions. Scaling analysis is firstly
discovered in physics and next it has been increasingly applied to
other disciplines especially in mathematics where theoretical
concepts are developed and extended. In financial studies, scaling
analysis have been applied for economic and/or financial data to
prove the existence of scaling laws such as financial prices,
exchange rates, etc. Scaling analysis of data gives useful
information on the process that generates such data. Such process
is the crucial point for understanding empirical and theoretical
mechanisms that generate the data and in using the empirical
scaling evidence as a stylized fact that any theoretical model
should also reproduce. Several models have been proposed in the
literature. However, they succeed in explaining some empirical
cases, but fail in many others. Wavelet methods initially
introduced in mathematics and physics have been applied in many
fields such as finance. We want in the present paper to provide
some contributions clarifying theses concepts, based on financial
data samples. We propose to reconstruct missing data parts in
order to obtain a complete basis. The completeness of missing data
bases is already evoked by many authors such as Aminghafari et al
(2007), Ben Mabrouk et al (2008), Karp (1988), Soltani (2002),
etc, and the studies in such a subject are increasing. However,
the classical methods such as Neural Networks, Fuzzy Logic,
Fourier Analysis, Autoregressive Models suffer from important
insufficiencies especially in causing noise components. Wavelet
methods have been proved to be the most powerful tool in
reconstructing missing parts already with eliminating the noise.
This is a first motivation in the application of wavelet analysis
in our study. A second motivation is due to the fact that wavelet
algorithms are usually most rapid yielding thus an important gain
in time and cost. Furthermore, it permits to overcome the non
disposability data problem and/or the hard data access.

One principal aim of the present work is to test the wavelet
method in constructing missing data yielding prediction prospects
and hence it may enable interesting functional regression
complements. One motivation of our idea is to provide a
longitudinal study to test the relation governance and
diversification strategies. Indeed, some tests conducted with
classical well known methods such as Neural Networks are developed
and yielded non efficient results. This resolves the cross-section
studies problem. Recall that the relation between diversification
strategies and performance has already been the object of several
studies such as Wernerfelt et al (1988), Amit et al (1988), etc.
However, even-though, there are enough studies on such a subject,
the results remains non concluding. See for instance Ramanujam et
al (1989). A detailed study on such a subject will be provided
later on complete sample data based on the theoretical results.

The present paper will be organized as follows. In the next
section, some basics on wavelet analysis are reviewed, essentially
those related to our application. In section 3, the proposed
prediction method of financial series and so time series is
developed based on wavelet estimators. It consists of a wavelet
series learning algorithm consisting in acting dynamic wavelet
estimators to build predicted values on short time series. The
main crucial point that makes the method important is its
simplicity in one hand especially for machine implementing and on
the other hand, because of the fact that it acts on governance,
performance and diversification variables. In such cases, the data
is characterized by short time horizons which may effect
negatively on empirical results. Indeed, in section 4, empirical
results are provided already with eventual discussions based on
comparisons with old methods such as Neural Networks. It is shown
that predicted values and also reconstructed values of real data
on Neural Networks are not adequate. We conclude afterwards.
\section{Wavelet analysis review}
Wavelet analysis has been practically introduced in the beginning
of the 80's in the context of signal analysis and oil exploration.
It gives a representation of signals permitting the simultaneously
enhancement of the temporal and the frequency information
(time-frequency localization). Its application comes to overcome
the insufficiencies of Fourier Analysis. It consists in
decomposing a series into different frequency components with a
scale adapted resolution. The advantage that it presents compared
to Fourier Analysis is the fact that it permits to observe and to
analyze data at different scales.Wavelets analysis proposed
initially by J. Morlet, is based on a concept somewhat different
from the frequency one: the concept of scale. Instead of
considering oscillating functions supported on a window, that can
be shifted along the support of the analyzed signal, wavelet basis
elements are copies of each other nearly compliant and only differ
by their sizes (Daubechies (1992), Gasquet and Witomsky (1990)).
Wavelet based methods are part of the most effective methods
currently used. The starting point in wavelet analysis is to
decompose a time series on scale-by-scale basis in order to
control the series structure at different horizons. A wavelet
basis is obtained from one source function $\psi$ known as mother
wavelet by dilation and translation operations. Each wavelet basis
element is defined for $j,k\in\mathbb{Z}$ as a copy of $\psi$ at
the scale $j$ and the position $k$,
$$
\psi_{j,k}(t)=2^{j/2}\psi(2^jt-k).
$$
The quantity $2^j$ designing the frequency of the series reflects
the dynamic behavior of the series according to the time variable.
This is why the index $j$ is usually called the frequency. From a
mathematical-physical point of view, the index $k$ localizes the
singularities of the series. In finance, or in data analysis the
index $k$ is used to localize the fluctuations and the missing
data periods. Let for $j\in\mathbb{Z}$ fixed, $W_j$ be the space
generated by $\left(\psi_{j,k}\right)_k$. Such a space is called
the $j$-level details space. A time series $X(t)$ is projected
onto $W_j$ yielding a component $X^d_j(t)$ given by
\begin{equation}\label{Wjprojection}
X_j^d(t)=\displaystyle\sum_kd_{j,k}\psi_{j,k}(t)
\end{equation}
where the $d_{j,k}$'s known as the wavelet or the detail
coefficients of the series $X(t)$ are obtained by
$$
d_{j,k}=<X;\psi_{j,k}>=\displaystyle\int_{\mathbb{R}}X(t)\psi_{j,k}(t)dt.
$$
We will see later the impact of these coefficients in financial
series analysis. In wavelet theory, the spaces $W_j$'s satisfy the
following orthogonal sum.
\begin{equation}\label{Wjsommedirecte}
\displaystyle\bigoplus_{j\in\mathbb{Z}}^{\bot}W_j=L^2(\mathbb{R}).
\end{equation}
This means that the series $X(t)$ can be completely reconstructed
via its detail components and that these components are mutually
uncorrelated. But a famous starting point in financial studies
before going on studying or detecting the instantaneous behavior
of the series, is to describe its global behavior or its trend.
This is already possible when applying wavelet analysis. Let us be
more precise and explain the idea. The mother wavelet yields a
second function called father wavelet or scaling function usually
denoted by $\varphi$. For more details on the relationship
$\psi/\varphi$ the readers can be referred to Daubechies (1992),
Gasquet (1990), Hardle et al (1996), etc. Similarly to $\psi$ the
scaling function $\varphi$ yields dilation-translation copies
$$
\varphi_{j,k}(t)=2^{j/2}\varphi(2^jt-k).
$$
Let $V_j$ be the space generated by
$\left(\varphi_{j,k}\right)_k$. Under suitable conditions on
$\psi$ or equivalently on $\varphi$ (Daubechies (1992)), the
sequence $\left(V_j\right)_j$ is called a multi-resolution
analysis (multi-scale analysis) on $\mathbb{R}$ and the $V_j$ is
called the $j$-level approximation space. It satisfies
\begin{equation}\label{VjincluVj+1}
V_j\subset\,V_{j+1},\quad\forall\,j\in\mathbb{Z}.
\end{equation}
This means that horizons $j$ and $j+1$ can be viewed from from
each other and so from any horizon $p\geq\,j+1$. In
physics-mathematics this is called the zooming characteristics of
wavelets. We have in fact a more precise relation of zooming,
\begin{equation}\label{zoom}
f\in\,V_j\Leftrightarrow\,f(2^j.)\in\,V_{0},\quad\forall\,j\in\mathbb{Z}.
\end{equation}
This reflects the fact that, not only the signal $f$ from horizon
$j$ can be seen in the horizon $j+1$ but either his contracted or
delated copies. We will see later its relation to financial data.
The sequence $\left(V_j\right)$'s satisfies also
\begin{equation}\label{intersection}
\overline{\displaystyle\bigcup_{j\in\mathbb{Z}}V_j}=L^2(\mathbb{R})
\quad\hbox{and}\quad
\displaystyle\bigcap_{j\in\mathbb{Z}}V_j=\{0\}.
\end{equation}
The first part means that it covers the whole space of finite
variance series. The second part means that there is no
correlations between the projections or the components of the time
series relatively to the different horizons. Finally the sequence
$\left(V_j\right)_j$ satisfies a shift invariance property in the
sense that
\begin{equation}\label{translation}
f\in\,V_j\Leftrightarrow\,f(.-k)\in\,V_{j},\quad\forall\,k\in\mathbb{Z}.
\end{equation}
This means that the multi-resolution analysis permits to detect
the properties of the signal as well as its shifted copies. i.e,
along the whole time support. Finally, the following relation
relates the $W_j$'s to the $V_j$'s,
\begin{equation}\label{Vj+Wj}
V_{j+1}=V_j\oplus\,W_j,\quad\forall\,j\in\mathbb{Z}.
\end{equation}
Under these properties, equations (\ref{Wjprojection}) and
(\ref{Wjsommedirecte}) yield, for $j\in\mathbb{Z}$ fixed, the
following decomposition
\begin{equation}\label{waveletseries}
X(t)=\displaystyle\sum_{j\in\mathbb{Z}}X_j^d(t)=\underbrace{\displaystyle\sum_{j\leq\,J-1}X_j^d(t)}_{X_J^a(t)}
+\displaystyle\sum_{j\geq\,J}X_j^d(t).
\end{equation}
Using equations (\ref{Wjsommedirecte}) and (\ref{Vj+Wj}), the
component $X^a_J(t)$ is the projection of the series $X(t)$ onto
the space $V_J$. Such a component is often called the
approximation of $X(t)$ at the level $J$. It represents the global
behavior of the series $X(t)$ or the trend at the scale $J$ while
the remaining part reflects the details of the series at different
scales. Using the definition of the spaces $V_j$'s, the component
$X^a_J(t)$ which belongs to $V_J$ can be expressed using the basis
$\left(\varphi_{J,k}\right)_k$ of such a space. Let
\begin{equation}\label{VJprojection}
X^a_J(t)=\displaystyle\sum_{k\in\mathbb{Z}}C_{J,k}^a\varphi_{J,k}(t)
\end{equation}
where the coefficients $C_{J,k}^a$, are obtained by
$$
C_{J,k}^a=<X;\varphi_{J,k}>=\displaystyle\int_{\mathbb{R}}X_J^a(t)\varphi_{J,k}(t)dt.
$$
It is proved in mathematical analysis that these did not depend on
$X^a_J$ but these are related directly to the original series
$X(t)$ by
$$
C_{J,k}^a=<X;\varphi_{J,k}>.
$$
So that, the dependence on the indexation $a$ will be omitted and
these will be denoted simply by $C_{J,k}$ and are often called the
scaling or the approximation coefficients of $X(t)$. The following
relation is obtained issued from (\ref{waveletseries}),
\begin{equation}\label{waveletseries1}
X(t)=\displaystyle\sum_{k\in\mathbb{Z}}C_{J,k}\varphi_{J,k}(t)
+\displaystyle\sum_{j\geq\,J}\displaystyle\sum_{k\in\mathbb{Z}}d_{j,k}\psi_{j,k}(t).
\end{equation}
This equality is known as the wavelet decomposition of $X(t)$. It
is composed of one part reflecting the global behavior or the
trend of the series and a second part reflecting the higher
frequency oscillations and so the fine scale deviations of the
trend.

To resume, in practice one can not obviously go to infinity in
computing the complete set of coefficients. So, we fix a maximal
level of decomposition $J_{max}$ and consider the decomposition
\begin{equation}\label{waveletdecompositionlevelJmax}
X_{J_{max}}(t)=\displaystyle\sum_{k\in\mathbb{Z}}C_{J,k}\varphi_{J,k}(t)
+\displaystyle\sum_{J\leq\,j\geq\,J_{max}}\displaystyle\sum_{k\in\mathbb{Z}}d_{j,k}\psi_{j,k}(t).
\end{equation}
There is no theoretical method for the exact choice of the
parameters $J$ and $J_{max}$. However, the minimal parameter $J$
does not have an important effect on the decomposition. But, the
choice of $J_{max}$ is always critical. One selects $J_{max}$
related to the error estimates.
\section{Methodology}
This section is devoted to present a wavelet based method to
reconstruct missing data. The method consists in providing a
prediction procedure able to predict a short time series on an
arbitrarily set of extending values. Recall that most of the
studies devoted to the prediction of time series are based on
three main ideas. Firstly, the disposed sample data is known on a
long time interval which is large enough than the predicted one.
Secondly, the learning procedure is conducted on the whole time
support which necessitates known data on this interval and thus a
reconstruction or a prediction will not be of importance while the
data is already known and it may be just for testing the accuracy
of methods. Finally, forecasting time series is based on the
disposition of a test sample. These facts are critical for many
reasons. Firstly, the availability of long samples is not already
possible. In contrast, for some situations such as the study of
the relationship ownership-structure/diversification in our case,
the samples are usually short. One has one main value on a year.
Secondly, one sometimes seeks to predict on a long period more
than the known one. Furthermore, when applying wavelet analysis to
approximate and/or to forecast time series, the majority of the
existing studies assume the presence of seasonality, periodicity
and/or autoregressive behavior in the series or in the wavelet
coefficients. See for instance Ben Mabrouk et al (2008), Soltani
(2002) and the references therein. In the present paper, we
provided a simple method leading to good prediction. The main
important point in our method as it will be shown later, is the
fact that it does not necessitate to test on the dynamic behavior
of the series and/or its wavelet coefficients. This is essentially
due to the fact that the considered series is short and its
dynamics may not be important. However, the most positive point in
the method is the fact that it necessitates only to compute the
values of the source scaling function and the associated wavelet
on a finite set of dyadic numbers.

Let $X(t)$ be a time series on a time domain
$\mathbb{T}=\{0,1,\dots,\,N\}$. A recursive procedure is used
which consists in applying firstly an estimator partially at short
horizons to all the observations $(t_i,X_i)_{i=1,\dots,N}$ to
yield firstly the predicted value $\widehat{X}_{N+1}$ of $X(N+1)$.
This last is then introduced as new observation to predict
$X_{N+2}$. We then follow the same steps until reaching the
desired horizon. Recall that the $J$-level wavelet decomposition
of the series $X(t)$, $t\in\mathbb{T}$, is
\begin{equation}\label{Jleveldecomposition}
X(t)=\underbrace{\displaystyle\sum_{k\in\mathbb{Z}}C_{0,k}\varphi_{0,k}(t)}_{X_{\varphi}(t)}
+\displaystyle\sum_{j=1}^J\underbrace{\displaystyle\sum_{k\in\mathbb{Z}}d_{j,k}\psi_{j,k}(t)}_{X_{j,\psi}(t)}.
\end{equation}
For $t=N+1$, this yields
\begin{equation}\label{X(N+1)}
X(N+1)=\displaystyle\sum_{k\in\mathbb{Z}}C_{0,k}\varphi(N+1-k)
+\displaystyle\sum_{j=1}^J\displaystyle\sum_{k\in\mathbb{Z}}d_{j,k}2^{j/2}\psi(2^j(N+1)-k).
\end{equation}
This means that for evaluating the predicted value
$\widehat{X}_{N+1}$, it suffices to do this for $X_{\varphi}$ and
the $X_{j,\psi}$'s. To attain this goal, it appears from
(\ref{X(N+1)}) that it suffices to compute the values of the
scaling function $\varphi$ and the wavelet mother $\psi$ on the
integer grid $\{N+1-k\,;\,k\in\mathbb{Z}\}$ and
$\{2^j(N+1)-k\,;\,k\in\mathbb{Z}\}$ on the supports of $\varphi$
and $\psi$. This motivates the use of Daubechies compactly
supported wavelets which are well evaluated on the integer grid.
(The values of the used scaling function and the wavelet
associated are provided in the appendix). Next,
$\widehat{X}_{N+1}$ is estimated as
\begin{equation}\label{Xchapeau(N+1)}
\widehat{X}_{N+1}=\widehat{X}_{\varphi}(N+1)+\displaystyle\sum_{j=1}^J\widehat{X}_{j,\psi}(N+1).
\end{equation}
where
\begin{equation}\label{Xchapeauphi}
\widehat{X}_{\varphi}(t)=\displaystyle\sum_{k\in\mathbb{Z}}\widehat{C}_{0,k}\varphi(t-k).
\end{equation}
and similarly
\begin{equation}\label{Xchapeaujpsi}
\widehat{X}_{j,\psi}(t)=\displaystyle\sum_{k\in\mathbb{Z}}\widehat{d}_{j,k}\varphi(t-k).
\end{equation}
with suitable estimators of the coefficients $\widehat{C}_{0,k}$
and $\widehat{d}_{j,k}$. The estimators are applied to avoid the
presence may be of high dynamics in the series of wavelet
coefficients and thus the possibility of noised parts and/or small
ones. Such estimators lead to $\widehat{X}_{N+1}$ which will be
added to the series and then we re-consider the initial step.
\section{Results, interpretations and discussions}
In this section, we develop empirical results based on the
theoretical procedure developed previously. We propose to apply
the step-by-step procedure described above to reproduce missing
data parts of the data basis composed of performance and
governance variables associated to French firms. The sample
studied is composed of governance, diversification and value
creation variables on a set of 69 French firms along the years
1995 to 2005. The wavelet method developed and Neural Networks are
tested in order to conduct eventual comparison to prove the
efficiency of our method. In fact, testing autoregressive
procedure is not efficient here due to short time interval. It is
known that for such cases the presence of autoregressive behavior
is weak and may not be proved. Fuzzy Logic is also not suitable
for the main reason of the absence of fuzzy data and the fact that
it necessitates to fix the rules which is based on expecting the
outputs which is ambiguous especially for governance,
diversification and value creation variables considered here.
Fourier Analysis necessitates the presence of periodicity which
can not be expected for our variables. This motivates the
comparison with Neural Networks. We just recall that we will not
expose all the results because of the largeness of the sample but
we restrict to some ones. Table 1 resumes the set of explicative
variables.
\begin{table}[ht]
\begin{center}
\begin{tabular}{||c|c||}
\hline
Variable&French Abbreviation\\
\hline
Total assets&AT\\
\hline
Market Capitalization&CB\\
\hline
Sales&CA\\
\hline
Total Equity&KP\\
\hline
Total Debt&DT\\
\hline
Net Income&Rnet\\
\hline
\end{tabular}
\caption{Analyzing Variables}\label{Table1}
\end{center}
\end{table}

The numerical results are provided using Daubechies compactly
supported wavelets by applying a $4$-level $DB10$ wavelet analysis
and a hard threshold wavelet estimator with a threshold
$\varepsilon=0.75Max|d_{j,k}|$. The neural network forecasts are
obtained by applying the well known software Alyuda Forecaster XL.
Such forecaster a part of input data as a training set to find the
best Neural Network and apply it by the next for the prediction.
One main disadvantage in it is the necessary hypothesis of
periodicity which may not be reel.
\subsection{Reconstruction of existing data}
Firstly, in order to test the efficiency and the performance of
the proposed idea we applied it for the known parts of our data
basis. A comparison is then conducted with reconstructed parts
using Neural Network method. The results are exposed in the
following tables where the effectiveness of the wavelet method
appears. For each variable and each firm we fix the 5 or 6 first
values to predict the 6 or 5 remaining ones. The tables 2-7 show
the predicted values of the different explicative variables for
the French firm ACCOR.
\begin{table}[ht]
\begin{center}
\begin{tabular}{||c|c|c|c||}
\hline
Year&Real\,\,Values&Wavelet\,\,Prediction&NN\,\,Prediction\\
\hline
1995&2799.22&\dots&\dots\\
\hline
1996&3146.46&\dots&\dots\\
\hline
1997&6086.85&\dots&\dots\\
\hline
1998&6504.18&\dots&\dots\\
\hline
1999&8707.07&\dots&\dots\\
\hline
2000&8820.38&8820.4&8694.861409\\
\hline
2001&8097.8&8097.8&8683.803108\\
\hline
2002&5733.19&5733.2&8672.965039\\
\hline
2003&7035.82&7035.8&8693.818931\\
\hline
2004&6455.4&6455.4&8699.095708\\
\hline
2005&9591.36&9591.4&8701.587759\\
\hline
\end{tabular}
\caption{Comparison results: CB for the French firm
Accor}\label{Table2}
\end{center}
\end{table}
\begin{table}[ht]
\begin{center}
\begin{tabular}{||c|c|c|c||}
\hline
Year&Real\,\,Values&Wavelet\,\,Prediction&NN\,\,Prediction\\
\hline
1995&3939.617175&\dots&\dots\\
\hline
1996&3482.397392&\dots&\dots\\
\hline
1997&3816.771372&\dots&\dots\\
\hline
1998&3262.152559&\dots&\dots\\
\hline
1999&4139.890318&\dots&\dots\\
\hline
2000&4305.977899&4306&3156.565015\\
\hline
2001&4087.818334&4087.8&4153.890745\\
\hline
2002&3812.589802&3812.6&3156.565015\\
\hline
2003&3816.340647&3816.3&4149.104216\\
\hline
2004&3839.705024&8339.7&3140.391532\\
\hline
2005&3828.989097&3829&4150.925829\\
\hline
\end{tabular}
\caption{Comparison results: DT for the French firm
Accor}\label{Table3}
\end{center}
\end{table}
\begin{table}[ht]
\begin{center}
\begin{tabular}{||c|c|c|c||}
\hline
Year&Real\,\,Values&Wavelet\,\,Prediction&NN\,\,Prediction\\
\hline
1995&8343.839611&\dots&\dots\\
\hline
1996&8450.858822&\dots&\dots\\
\hline
1997&9725.027708&\dots&\dots\\
\hline
1998&9423.636001&\dots&\dots\\
\hline
1999&10865&\dots&\dots\\
\hline
2000&11954&11954&9410.595452\\
\hline
2001&12100&12100&8182.470332\\
\hline
2002&11275&11275&9317.530091\\
\hline
2003&10956&10956&8073.268155\\
\hline
2004&11510&11510&9342.956958\\
\hline
2005&12791&12791&8262.130542\\
\hline
\end{tabular}
\caption{Comparison results: AT for the French firm
Accor}\label{Table4}
\end{center}
\end{table}
\begin{table}[ht]
\begin{center}
\begin{tabular}{||c|c|c|c||}
\hline
Year&Real\,\,Values&Wavelet\,\,Prediction&NN\,\,Prediction\\
\hline
1995&2025.132745&\dots&\dots\\
\hline
1996&2459.764893&\dots&\dots\\
\hline
1997&2843.783968&\dots&\dots\\
\hline
1998&2873.968873&\dots&\dots\\
\hline
1999&3092&\dots&\dots\\
\hline
2000&3843&3843&3091.741598\\
\hline
2001&4139&4139&3091.735656\\
\hline
2002&3893&3893&309173147\\
\hline
2003&3587&3587&3090.331298\\
\hline
2004&3755&3755&3088.660429\\
\hline
2005&4301&4301&3088.112759\\
\hline
\end{tabular}
\caption{Comparison results: KP for the French firm
Accor}\label{Table5}
\end{center}
\end{table}
\begin{table}[ht]
\begin{center}
\begin{tabular}{||c|c|c|c||}
\hline
Year&Real\,\,Values&Wavelet\,\,Prediction&NN\,\,Prediction\\
\hline
1995&4605.789709&\dots&\dots\\
\hline
1996&4251.498193&\dots&\dots\\
\hline
1997&4781.563426&\dots&\dots\\
\hline
1998&5554.175045&\dots&\dots\\
\hline
1999&6044&\dots&\dots\\
\hline
2000&6946&6946&6036.620238\\
\hline
2001&7218&7218&6034.605972\\
\hline
2002&7071&7071&6034.053505\\
\hline
2003&6774&6774&6043.649803\\
\hline
2004&7072&7072&6041.789193\\
\hline
2005&7562&7562&6040.57782\\
\hline
\end{tabular}
\caption{Comparison results: CA for the French firm
Accor}\label{Table6}
\end{center}
\end{table}
\begin{table}[ht]
\begin{center}
\begin{tabular}{||c|c|c|c||}
\hline
Year&Real\,\,Values&Wavelet\,\,Prediction&NN\,\,Prediction\\
\hline
1995&44.89044071&\dots&\dots\\
\hline
1996&163.7937425&\dots&\dots\\
\hline
1997&185.2322648&\dots&\dots\\
\hline
1998&259.4859474&\dots&\dots\\
\hline
1999&267.7294338&\dots&\dots\\
\hline
2000&431.8525675&431.8526&267.5776237\\
\hline
2001&473.6313588&473.6314&267.5647551\\
\hline
2002&372.713343&372.7133&267.6051109\\
\hline
2003&267.9924661&267.9925&267.6925082\\
\hline
2004&242.3262479&242.3262&267.6993183\\
\hline
2005&331.61336&331.6134&267.6998488\\
\hline
\end{tabular}
\caption{Comparison results: Rnet for the French firm
Accor}\label{Table7}
\end{center}
\end{table}
\newpage
\subsection{Construction of missing data}
As it is shown in tables 2-7, the wavelet predicted values (WPV)
are more accurate compared to neural networks predicted ones
(NNPV). This motivates the application of the wavelet procedure to
reproduce the really missing parts of our series taking in mind
that the obtained values will be the best. The empirical tests are
conducted on the French firm DMC for which we do not dispose of
the data for the years 2000 to 2005. The results are exposed in
tables 8-13.
\begin{table}[ht]
\begin{center}
\begin{tabular}{||c|c|c|c||}
\hline
Year&Real\,\,Values&Wavelet\,\,Prediction&NN\,\,Prediction\\
\hline
1995&198.76&\dots&\dots\\
\hline
1996&134.61&\dots&\dots\\
\hline
1997&127.02&\dots&\dots\\
\hline
1998&77.35&\dots&\dots\\
\hline
1999&45.75&\dots&\dots\\
\hline
2000&Unknown&101.32&45.42349633\\
\hline
2001&Unknown&116.75&45.42349633\\
\hline
2002&Unknown&74.04&45.42349633\\
\hline
2003&Unknown&83.4&46.39115105\\
\hline
2004&Unknown&61.42&46.18757939\\
\hline
2005&Unknown&57.36&46.2256794\\
\hline
\end{tabular}
\caption{Comparison results: CB for the French firm
DMC}\label{Table8}
\end{center}
\end{table}
\begin{table}[ht]
\begin{center}
\begin{tabular}{||c|c|c|c||}
\hline
Year&Real\,\,Values&Wavelet\,\,Prediction&NN\,\,Prediction\\
\hline
1995&140.1223053&\dots&\dots\\
\hline
1996&151.4085821&\dots&\dots\\
\hline
1997&136.6443018&\dots&\dots\\
\hline
1998&236.3393133&\dots&\dots\\
\hline
1999&203.168443&\dots&\dots\\
\hline
2000&Unknown&217.0818&217.0817999\\
\hline
2001&Unknown&124.8029&124.8028628\\
\hline
2002&Unknown&95.9582&95.95824202\\
\hline
2003&Unknown&85.8677&85.86766452\\
\hline
2004&Unknown&68.9464&68.94638019\\
\hline
2005&Unknown&76.0819&76.08186408\\
\hline
\end{tabular}
\caption{Comparison results: DT for the French firm
DMC}\label{Table9}
\end{center}
\end{table}
\begin{table}[ht]
\begin{center}
\begin{tabular}{||c|c|c|c||}
\hline
Year&Real\,\,Values&Wavelet\,\,Prediction&NN\,\,Prediction\\
\hline
1995&801.119586&\dots&\dots\\
\hline
1996&739.98753&\dots&\dots\\
\hline
1997&660.409143&\dots&\dots\\
\hline
1998&643.029955&\dots&\dots\\
\hline
1999&416.795613&\dots&\dots\\
\hline
2000&Unknown&339.1&415.8189638\\
\hline
2001&Unknown&236.7&415.8189638\\
\hline
2002&Unknown&178.3&415.8189638\\
\hline
2003&Unknown&163.4&421.4230757\\
\hline
2004&Unknown&120.6&435.9540594\\
\hline
2005&Unknown&120.6&496.7405454\\
\hline
\end{tabular}
\caption{Comparison results: AT for the French firm
DMC}\label{Table10}
\end{center}
\end{table}
\begin{table}[ht]
\begin{center}
\begin{tabular}{||c|c|c|c||}
\hline
Year&Real\,\,Values&Wavelet\,\,Prediction&NN\,\,Prediction\\
\hline
1995&355.053761&\dots&\dots\\
\hline
1996&276.694966&\dots&\dots\\
\hline
1997&221.660871&\dots&\dots\\
\hline
1998&131.563502&\dots&\dots\\
\hline
1999&30.794701&\dots&\dots\\
\hline
2000&Unknown&511.3528&29.39673892\\
\hline
2001&Unknown&482.7281&29.39673892\\
\hline
2002&Unknown&503.9588&29.39673892\\
\hline
2003&Unknown&568.4346&32.31103077\\
\hline
2004&Unknown&392.2119&36.16726538\\
\hline
2005&Unknown&189.8276&42.57971843\\
\hline
\end{tabular}
\caption{Comparison results: KP for the French firm
DMC}\label{Table11}
\end{center}
\end{table}
\begin{table}[ht]
\begin{center}
\begin{tabular}{||c|c|c|c||}
\hline
Year&Real\,\,Values&Wavelet\,\,Prediction&NN\,\,Prediction\\
\hline
1995&1085.437003&\dots&\dots\\
\hline
1996&948.690234&\dots&\dots\\
\hline
1997&906.461856&\dots&\dots\\
\hline
1998&818.041426&\dots&\dots\\
\hline
1999&667.879145&\dots&\dots\\
\hline
2000&Unknown&522.9&669.1731789\\
\hline
2001&Unknown&384.2&669.6223291\\
\hline
2002&Unknown&307.4&669.8566586\\
\hline
2003&Unknown&250.7&672.4642615\\
\hline
2004&Unknown&206&676.8128998\\
\hline
2005&Unknown&187.3&684.7212645\\
\hline
\end{tabular}
\caption{Comparison results: CA for the French firm
DMC}\label{Table12}
\end{center}
\end{table}
\begin{table}[ht]
\begin{center}
\begin{tabular}{||c|c|c|c||}
\hline
Year&Real\,\,Values&Wavelet\,\,Prediction&NN\,\,Prediction\\
\hline
1995&0.160322851&\dots&\dots\\
\hline
1996&-91.65018436&\dots&\dots\\
\hline
1997&-73.25569807&\dots&\dots\\
\hline
1998&-89.23636453&\dots&\dots\\
\hline
1999&-101.2818874&\dots&\dots\\
\hline
2000&Unknown&-82.9199&-102.3564978\\
\hline
2001&Unknown&-23.3818&-102.6829817\\
\hline
2002&Unknown&-3.7065&-102.6829817\\
\hline
2003&Unknown&8.1046&-102.5151181\\
\hline
2004&Unknown&-8.1113&-102.4520249\\
\hline
2005&Unknown&-11.7509&102.4326179\\
\hline
\end{tabular}
\caption{Comparison results: Rnet for the French firm
DMC}\label{Table13}
\end{center}
\end{table}
\subsection{Performance and Value Creation Variables}
The aim of this section is to inject the wavelet predicted or
reconstructed variables into the equations yielding the endogenous
variables. Two methods will be exposed. First, we provide the
values of performance and value creation variables as endogenous
variables, based on their explicit dependencies on the explicative
variables reconstructed in the previous subsections. Secondly, we
provide wavelet predictions of the endogenous variables using the
wavelet procedure developed in section 2. A comparison between
these values is also conducted as well as with a comparison with
existing values for DMC firm and with Neural Network predicted
values for the two firms. The performance variables and value
creation ones constituting the endogenous variables are listed in
table 14 with their explicit expressions on the explicative ones.
The empirical tests on DMC firm are provided in tables 15-18.
\begin{table}[ht]
\begin{center}
\begin{tabular}{||c|c|c||}
\hline
Variables&Abbreviations&Explicitness\\
\hline
Economic performance&Q&(CB+DT)/AT\\
\hline
Value creation&RMRS&CB/KP\\
\hline
Return on equity&ROE&Rnet/KP\\
\hline
Return on assets&ROA&Rnet/AT\\
\hline
\end{tabular}
\caption{Performance and value creation variables}\label{Table14}
\end{center}
\end{table}
\begin{table}[ht]
\begin{center}
\begin{tabular}{||c|c|c|c||}
\hline
Year&Real\,\,Values&Wavelet\,\,Predicted\,\,Values&Explicit\,\,Reconstruction\,\,with\,\,WPV\\
\hline
1995&0.807642223&\dots&\dots\\
\hline
1996&0.784400442&\dots&\dots\\
\hline
1997&1.018364335&\dots&\dots\\
\hline
1998&1.03636564&\dots&\dots\\
\hline
1999&1.182416964&\dots&\dots\\
\hline
2000&1.098072436&1.0981&1.0981\\
\hline
2001&1.007075895&1.0071&1.0071\\
\hline
2002&0.846632355&0.8466&0.8466\\
\hline
2003&0.990522147&0.9905&0.9905\\
\hline
2004&0.894448742&0.8944&0.8944\\
\hline
2005&1.049202494&1.0492&1.0492\\
\hline
\end{tabular}
\caption{Comparison results: Comparison results: Tobin's Q for the
French firm DMC}\label{Table15}
\end{center}
\end{table}
\begin{table}[ht]
\begin{center}
\begin{tabular}{||c|c|c|c||}
\hline
Year&Real\,\,Values&Wavelet\,\,Predicted\,\,Values&Explicit\,\,Reconstruction\,\,with\,\,WPV\\
\hline
1995&1.382240254&\dots&\dots\\
\hline
1996&1.279171033&\dots&\dots\\
\hline
1997&2.140405203&\dots&\dots\\
\hline
1998&2.263135158&\dots&\dots\\
\hline
1999&2.815999353&\dots&\dots\\
\hline
2000&2.295180848&2.2952&2.2952\\
\hline
2001&1.956462914&1.9565&1.9565\\
\hline
2002&1.472692011&1.4727&1.4727\\
\hline
2003&1.961477558&1.9615&1.9615\\
\hline
2004&1.719147803&1.7191&1.7191\\
\hline
2005&2.230030226&2.2300&2.2300\\
\hline
\end{tabular}
\caption{Comparison results: Comparison results: RMRS for the
French firm DMC}\label{Table16}
\end{center}
\end{table}
\begin{table}[ht]
\begin{center}
\begin{tabular}{||c|c|c|c||}
\hline
Year&Real\,\,Values&Wavelet\,\,Predicted\,\,Values&Explicit\,\,Reconstruction\,\,with\,\,WPV\\
\hline
1995&0.06948208&\dots&\dots\\
\hline
1996&0.06557174&\dots&\dots\\
\hline
1997&0.08084057&\dots&\dots\\
\hline
1998&0.10349035&\dots&\dots\\
\hline
1999&0.11384217&\dots&\dots\\
\hline
2000&0.11631538&0.1163&0.1124\\
\hline
2001&0.11452042&0.1145&0.1144\\
\hline
2002&0.11045466&0.1105&0.0957\\
\hline
2003&0.07527181&0.0753&0.0772\\
\hline
2004&0.06364847&0.0636&0.0645\\
\hline
2005&0.07742385&0.0774&0.0771\\
\hline
\end{tabular}
\caption{Comparison results: Comparison results: ROE for the
French firm DMC}\label{Table17}
\end{center}
\end{table}
\begin{table}[ht]
\begin{center}
\begin{tabular}{||c|c|c|c||}
\hline
Year&Real\,\,Values&Wavelet\,\,Predicted\,\,Values&Explicit\,\,Reconstruction\,\,with\,\,WPV\\
\hline
1995&0.00538007&\dots&\dots\\
\hline
1996&0.019381905&\dots&\dots\\
\hline
1997&0.019046965&\dots&\dots\\
\hline
1998&0.02753565&\dots&\dots\\
\hline
1999&0.024641457&\dots&\dots\\
\hline
2000&0.036126198&0.0361&0.0361\\
\hline
2001&0.039143088&0.0391&0.0391\\
\hline
2002&0.033056616&0.0331&0.0331\\
\hline
2003&0.025282262&0.0253&0.0253\\
\hline
2004&0.02105354&0.0211&0.0211\\
\hline
2005&0.025925523&0.0259&0.0259\\
\hline
\end{tabular}
\caption{Comparison results: Comparison results: ROA for the
French firm DMC}\label{Table18}
\end{center}
\end{table}
\section{Conclusion}
The study of the relationship ownership-structure/diversification
is of great interest essentially for two main reasons. Firstly,
the diversification constitutes a strategic important choice which
strongly affects on the capacity of firm development during a long
period, its financing needs, its performance as well as the
related risks. Secondly, such a decision may cause some
inter-actors interest conflicts in the firm and it may be imposed
within its actionnarial structure. The main problem in studying
the relationship is the necessity of conducting it on a complete
data basis. So, when having a missing data sample, the study can
not be well conducted and no good conclusions can be pointed out.
This motivates our work here based of constructing the missing
data parts in order to obtain a complete sample to be applied for
testing the ownership structure and diversification relationship.
As it is well known nowadays, wavelet theory is the most powerful
tool for filling this gap, we provided in the present work a
wavelet based method to complete the missing parts of an
incomplete sample composed of governance, diversification and
value creation variables on a set of 69 French firms along the
years 1995 to 2005. Some existing parts are also reconstructed in
order to test the efficiency of our method. Having now a complete
sample, we intend in an extending forthcoming work to apply the
obtained full sample for the study of the original object, the
relation between the properties of the structure and the
diversification. We intend as well to join the determinants and
the consequences of the diversification strategy.
\section{Appendix}
We refer to \cite{BenMohamed} for the computation of the values of
Daubechies father and mother wavelets. Recall that DB10 is
supported on the interval $[0,19]$. The grids $\varphi(n)$ and
$\psi(n)$ for $n$ integer in the support, are given in the
following table. The values on the whose dyadic grid are obtained
obviously using the 2-scale relation.
\begin{table}[ht]
\begin{center}
\begin{tabular}{||c|c|c||}
\hline
n&$\varphi(n)$&$\psi(n)$\\
\hline
1&3.354408256841158E-002&-1.668296029790936E-005\\
\hline
2&0.652680627783427&-1.413114882338799E-004\\
\hline
3&0.555223194777502&2.348833526687438E-003\\
\hline
4&-0.380687440933945&-1.203933191141380E-002\\
\hline
5&0.202266079588952&3.439149845726070E-002\\
\hline
6&-8.039450480025792E-002&-6.490697134847506E-002\\
\hline
7&1.740357229364825E-002&9.676861895219263E-002\\
\hline
8&1.788811154355532E-003&-0.176355684599155\\
\hline
9&-2.262291980513165E-003&0.563213124163635\\
\hline
10&3.861859807090320E-004&-0.658797645066142\\
\hline
11&7.746490117092401E-005&-0.359446985391733\\
\hline \hline
12&-2.595735132421100E-005&-3.947780645895724E-002\\
\hline \hline
13&7.455766803700000E-008&1.860748282571044E-003\\
\hline \hline
14&1.064649412020000E-007&1.050025200672993E-004\\
\hline \hline
15&-5.018348300000000E-009&-9.779523219362094E-006\\
\hline \hline
16&1.350623000000000E-011&2.927524652329252E-008\\
\hline \hline
17&9.591699999999999E-014&1.928006355576058E-010\\
\hline \hline
18&8.000000000000001E-018&-3.615587627311547E-015\\
\hline
\end{tabular}
\caption{Values of Daubechies 10, $\varphi$ and
$\psi$}\label{Tabledb10}
\end{center}
\end{table}

\end{document}